%% file: ms.tex
\documentclass[letterpaper,twocolumn,10pt]{article}
\usepackage{usenix,epsfig}

\usepackage[justification=justified,singlelinecheck=false]{caption}

\usepackage[hidelinks]{hyperref}
\makeatletter
\g@addto@macro{\UrlBreaks}{\UrlOrds}
\makeatother

\usepackage{graphicx}

\usepackage{multirow}

\usepackage{adjustbox}
\usepackage{array}

\newcolumntype{R}[2]{%
    >{\adjustbox{angle=#1,lap=\width-(#2)}\bgroup}%
    l%
    <{\egroup}%
}

\newcommand*\rot{\multicolumn{1}{R{45}{1em}}}

\usepackage{wasysym}

\newcolumntype{C}[1]{>{\centering\let\newline\\\arraybackslash\hspace{0pt}}m{#1}}

\usepackage{color}

\newcommand{\junk}[1]{}

\newcommand{\anonymized}[2]{{#2}}

\begin{document}
\date{}

\title{\vspace{-2em}\Large \bf The Future of Ad Blocking:\\ An Analytical Framework and New Techniques}

\author{
{\rm Grant Storey}\\
Princeton University
\and
{\rm Dillon Reisman}\thanks{Dillon Reisman is an independent contractor working with the Princeton Web Transparency and Accountability Project.}\\
Princeton University
\and
{\rm Jonathan Mayer}\\
Stanford University
\and
{\rm Arvind Narayanan}\\
Princeton University
} 

\maketitle



\subsection*{Abstract}
We present a systematic study of ad blocking --- and the associated ``arms race'' --- as a security problem. We model ad blocking as a state space with four states and six state transitions, which correspond to techniques that can be deployed by either publishers or ad blockers. We argue that this is a complete model of the system. We propose several new ad blocking techniques, including ones that borrow ideas from rootkits to prevent detection by anti-ad blocking scripts. Another technique uses the insight that ads must be recognizable by humans to comply with laws and industry self-regulation. We have built prototype implementations of three of these techniques, successfully blocking ads and evading detection.

We systematically evaluate our proposed techniques, along with existing ones, in terms of security, practicality, and legality. We characterize the order of growth of the {\em development effort} required to create/maintain ad blockers as a function of the growth of the web. Based on our state-space model, our new techniques, and this systematization, we offer insights into the likely ``end game'' of the arms race.  We challenge the widespread assumption that  the arms race will escalate indefinitely, and instead identify a combination of evolving technical and legal factors that will determine the outcome.

\input{intro}

\input{related}

\input{model}

\input{newTechniques}

\input{systematization}

\input{legal}

\section{Conclusion}

In this paper we have presented an approach to ad blocking which is radically different from current techniques. Current ad blocking is based on the laborious process of creating filter rules and is easily disrupted by obfuscation implemented by publishers. In contrast we take a principled approach to the problem and present solutions that are easier to implement and harder to evade. 
Our work refutes the belief that the battle between publishers and users is leading to a permanent arms race between the two parties, and presents a much more nuanced picture.

Our rootkit technique for stealthy modifications to the browser may be of independent interest. In particular, ad fraud detection relies heavily on detecting differences in rendering behavior between legitimate and fraudulent browsers. Thus, stealth can be used as an attack on ad fraud detection, necessitating research into countermeasures.

Ad blocking is an important area of study for the security community. It combines old techniques from the domains of malware and program analysis with some new concepts such a mimicking human behavior. There is a significant need for follow-on technical work into the expansion of techniques that we have introduced as well as a debate on the ethics of ad blocking.

{\bf Acknowledgment.} We are grateful to Steven Englehardt and Harry Kalodner for feedback on the paper. Narayanan and Reisman are funded by NSF Award CNS 1526353.

{\footnotesize \bibliographystyle{acm}
\bibliography{biblio}

\input{appendices}

\end{document}

%% file: intro.tex
\section{Introduction}

\anonymized{Online advertising is a major part of the modern web, allowing websites to make money without direct payments from users. In recent years, however, there has been a rise in use of ad blockers. In response to this threat to revenue, advertisers and publishers have sought to circumvent ad blockers or prevent ad blocking users from viewing their sites, leading to an ``arms race'' as the two sides attempt to gain an advantage over each other. We study this arms race as a security problem.}{We study ad blocking - and the associated ``arms race'' - as a security problem.} Two reasons motivate this view. First, online ads and their supporting infrastructure create security threats such as malvertising. Thus, ad blocking is a security-enhancing measure. Second, ad blockers and anti-adblocking scripts can be seen as mutually hostile pieces of code executing in a shared environment. This scenario closely resembles traditional security settings such as malware vs anti-malware tools, and hence well-known security techniques can be adapted for ad blocking.

Most previous discussions of ad blocking have assumed that an arms race will escalate indefinitely \cite{register-arms-race,chicago-tribune-arms-race} or that it will favor publishers/advertisers rather than users \cite{wired-arms-race}. We challenge this view. Our argument is based on two key observations. First, ads must be recognizable by {\em humans} due to legal requirements imposed on online advertising (Section \ref{sec:techniques_perceptual}). Thus we propose {\em perceptual ad blocking} which works radically differently from current ad blockers. It deliberately ignores useful information in markup and limits itself to visually salient information, mimicking how a human user would recognize ads. We use lightweight computer vision techniques to implement such a tool and show that it defeats attempts to obfuscate the presence of ads.

Our second key observation is that even though publishers increasingly deploy scripts to detect and disable ad blocking, ad blockers run at a higher privilege level than such scripts, and hence have the upper hand in this arms race. We borrow ideas from rootkits to build a stealthy adblocker that evades detection. Our approach to hiding the presence and purpose of a browser extension is general and might be of independent interest. 

Specifically, we make four contributions. First, we model the ``state space'' of actions that can be taken by users seeking to block ads and by websites seeking to compel users to view ads. The space has four states, with two that represent ``success'' for the user and two for the website. We argue that this set of four states is a {\em complete} model of the system. In contrast to the intuition of eternal escalation, we show that it breaks down into a set of three ``mini-arms races'' which are comparatively easy to analyze. 

Second, we design and prototype several  client-side tools that demonstrate key capabilities that users (or ad blockers) possess in the ongoing ad blocking wars.\footnote{The code for all of our prototypes is available \anonymized{in anonymized form at \url{https://github.com/futureAdblockingPaper/ad-blocking}}{at \url{https://github.com/citp/ad-blocking}}.} The first, perceptual ad blocking, demonstrates users' ability to block ads despite websites' attempts to make such ads indistinguishable from other content. We released a prototype tool in mid-April 2017, which has about 30,000 users on the Chrome web store as of this writing.

The second technique, stealth, is a response to ``ad blocker blockers'' --- recently deployed on several sites --- that check for correct loading of ads.  The final technique, {\em active ad blocking}, replaces stealth with offense: it aims to identify and disable anti-adblocking JavaScript code. 

Third, we introduce a new systematization for evaluating ad blockers, whether existing tools, our new prototype tools, or future tools. It reveals that ad blockers must satisfy a complex set of sometimes-conflicting requirements encompassing security, deployability, and development effort. The labor-intensive nature of these tools has often been neglected in the discussion. We introduce a way to characterize the order of growth of the {\em programming effort} of different ad blocking tools in terms of the scale of the web.

Fourth, we explain that ad blocking implicates a range of potential sources of legal liability, including contract law, copyright law, the Digital Millennium Copyright Act, and the Computer Fraud and Abuse Act. We analyze the ad blocking techniques that we describe under each of these sources of law.
\anonymized{\footnote{The authors of this paper include a legal scholar.}}{}
While the landscape is complex and largely untested in court, we use legal reasoning grounded in past cases to conclude that most of the novel techniques can be architected in a way that minimizes legal risk. The details of our analysis are US-centric but the broad conclusions should apply to other jurisdictions as well. 

The techniques we propose are not perfect, but they are principled. They give us evidence that if the arms race continues, users have reasons to be optimistic. A favorable legal climate and the existence of browsers friendly towards ad blocking extensions are two key factors that may tip the scales towards users. Our work serves as a case study in the interplay between legal constraints and security techniques. Unlike the behavior of malware, the behavior of both publishers/advertisers and adblocking tools already is, and will continue to be, shaped by regulations. Finally, our work gives further urgency to the debate about the ethics of ad blocking and the financial future of web publishing.

%% file: related.tex
\section{Related Work} \label{sec:related_work}

Users block ads when browsing the web to minimize annoyance and ``creepiness'' \cite{cranor-creepy,turow-reject}, to improve performance \cite{kontaxis2015tracking}, and due to security concerns \cite{dwyermalvertising}. A number of studies have measured the prevalence of ad blocking. Analysis of 2015 Neilsen survey data from 11 countries reported a rate of 26\% of devices blocking ads \cite{young2016meet}. A measurement study using 2015 data from a European ISP reported a 22\% usage rate among the most active users \cite{pujol2015annoyed}. A previous study using 2012--2014 ISP data reported a rate of 10\%--18\% of households with at least one device having Adblock Plus installed \cite{metwalley2015online}. PageFair's annual report, which combines multiple data sources such as ad blocker / filter list downloads and website analytics, reports a worldwide rate of 11\% and North America rate of 18\% for 2016 \cite{pagefair-adblock-report}. 

Numerous authors have attempted to develop techniques for ad blocking which reduce the manual effort of assembling a massive filter list. Researchers have suggested the use of machine learning to enhance filter lists \cite{gugelmann2015automated}. Crowdsourcing-based techniques seek to dynamically develop filter lists based off of users' labeling of advertisements \cite{esfandiari2005adaptive, critesautomatic}. Many studies use automated (especially machine learning) approaches for detection of trackers; it is unclear to what extent these could apply to ads \cite{bau2013promising,yu2016tracking,ikram2017towards}.


A number of efforts have been made to evaluate the effectiveness of current ad blockers \cite{pujol2015annoyed, wills2016ad, singh2009blocking, mayer2012third}.
As ad blockers have emerged on the web, work has gone into detecting them and blocking users who make use of them \cite{iab2016}. As usage of these techniques has increased, they have been formalized and measured \cite{mughees2016first, nithyanand2016ad}. Continuing the arms race, tools are being developed to block ad blocking detectors themselves \cite{antiadblockkiller}.

This arms race has parallels with ad fraud techniques and scripts that seek to detect them. Ad fraud tools seek to simulate a real user browsing the web and viewing / clicking on ads. To be effective, they must spoof indicators such as whether an ad is viewable on the screen, and hide their presence from detection scripts (or disable those scripts). A whitepaper by anti-ad fraud firm White Ops discusses an ad fraud tool known as Methbot \cite{methbot}. While details are scant, Methbot appears to have developed techniques similar to ours, but arguably more rudimentary. This technical convergence is noteworthy in itself,  but also highlights the ethical concerns around this type of research, as our techniques might be misused in the ad fraud domain.



%% file: model.tex
\section{Model and state space} \label{sec:model}

\subsection{Model}
We aim to analyze the ``arms race'' between users and publishers: users seek to hide ads from web pages, but otherwise leave content and functionality unaffected. Publishers, to a first approximation, seek to ensure that users can view content or interact with functionality if and only if they view ads. To formulate this as a security problem, we need to concretely state a number of intuitive assumptions.

\textbf{The actors and their software}. We assume that the browser (and, if necessary, the device and operating system) is completely under the control of the user and modifiable as desired. We model ad blocking tools not as independent agents but rather as software tools acting on behalf of the user. Similarly, the publisher can modify all first-party content and code as well as third-party scripts as desired.

This assumption has some limitations in practice. Browser vendors who are funded by advertising may not be sympathetic to ad blocking. Other browsers, especially on mobile devices, may be ``locked down'' to varying degrees. We recognize these limitations, but argue that at least {\em some} browser vendors will allow ad blockers to exist and flourish, provided user demand continues to be strong. 
Indeed,  Safari on iOS recently modified its policy and now allows ad blocking \cite{ios-ad-blocking} and  Firefox added tracking protection features natively even though these are also opposed by the online advertising industry~\cite{moz-tracking-protection}. Opera has also recently added native ad blocking to speed up users' browsing experiences \cite{opera-native-blocking}. Finally, the new Brave browser has a business model that centers around ad blocking as a built-in feature. We emphasize that our focus is on the long-term evolution of ad blocking. 

\textbf{Legal boundaries and perceptual ad blocking.} 
Publishers and advertisers must abide by legal prohibitions against misleading advertising. In the United States, advertising disclosures are primarily regulated by the Federal Trade Commission (FTC). The FTC is empowered by Congress to take action against ``unfair'' and ``deceptive'' business practices, and the agency has adopted the position in guidelines and enforcement actions that paid advertisements must be clearly recognizable to consumers as such \cite{ftc-1, ftc-2}. An advertiser that relies on misleading content has primary liability under the FTC Act; a publisher could face secondary liability for enabling the misleading advertising. The European Union has adopted similar disclosure requirements for online advertising, in both the E-Commerce Directive (2000/31/EC) and the Unfair Commercial Practices Directive (2005/29/EC).

\begin{figure}
  \begin{center}
  \includegraphics{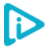}
  \caption{The AdChoices icon. AdChoices is an industry standard for disclosure of online behavioral advertising.}
  \label{fig:adchoices}
  \end{center}
\end{figure}

Online {\em behavioral} advertising, which includes much of online advertising today, is further governed by industry self-regulation. As a component of self-regulation in both North America and Europe, behavioral advertisements are usually labeled with a distinctive ``AdChoices'' icon (Figure~\ref{fig:adchoices}).

The requirement for ads to be recognizable by humans is at the core of one of our contributions in this paper. We introduce the notion of {\em perceptual ad blocking}, where the ad blocker deliberately limits itself to visually salient information, and recognizes ads just as a human would.

In addition to laws, market forces also have a role to play. We note that publishers could adopt a range of advertising techniques that disrupt the user experience, such as forcing a user to watch a video advertisement and answer questions based on it before allowing access to content. While these intrusive designs are technically trivial, competition among publishers has so far prevented such adversarial attempts at monetization.

\textbf{Blocking tracking is out of scope.} In current practice, blocking of ads and blocking of third-party trackers have been closely integrated, and have been seen as instances of the same problem. We argue for a different view. Ensuring that an ad is not seen by the user is (or can be) a purely client-side intervention, whereas blocking tracking requires preventing some network requests from being made. Perceptual ad blocking, which we introduce, is not capable of blocking trackers, since trackers are invisible to users. Conversely, users might choose to block tracking (say, via Tor browser) while allowing ads. Thus, the two problems are distinct, and we do not consider tracker blocking in this paper.


\textbf{Long-term outlook.} Our goal is to analyze the long-term evolution of ad blocking. Thus, in evaluating different designs, performance is only a minor criterion  since browser performance has tended to improve greatly over the years. Our design space includes ideas such as shadow execution of the entire page. Similarly, we consider designs that are not implementable in a extension in mainstream browsers today, provided that they can be implemented by a source-code modification. As noted earlier, there is at least one startup that aims to ship a browser with ad blocking built in \cite{brave}.

\subsection{State space}

\begin{figure}
  \centering
  \includegraphics[width=0.8\linewidth]{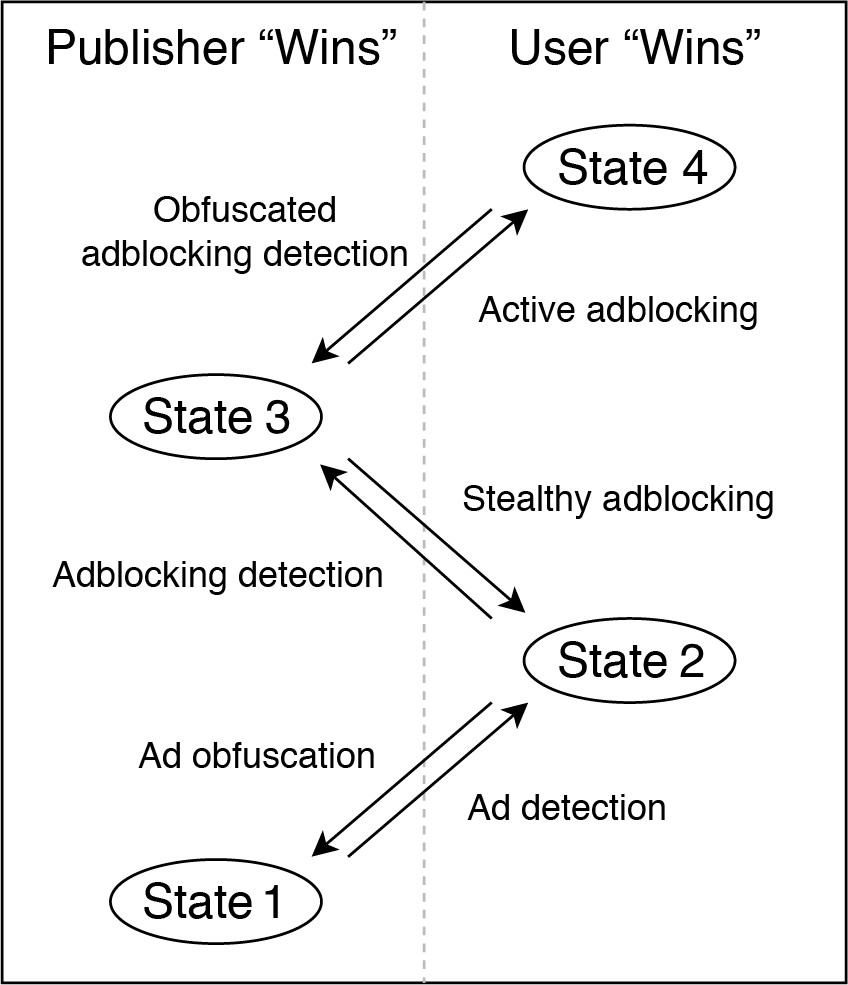}
  \caption{The state space. Our model reveals the structure behind the ad blocking ``arms race.'' Note that our model combines publishers and advertisers, and similarly users and ad blockers.}
  \label{fig:statespace}
\end{figure}

The state diagram in Figure \ref{fig:statespace} is the foundation of our contributions. A {\em(user, publisher)} pair is in one of these states at any point in time. A user may be in different states with respect to different publishers (in reality the granularity may be even finer, with some ads on a page being blocked but not others, but this is not a useful distinction for us). State transitions happen when the user employs an ad blocking technique or the publisher employs an anti-ad blocking technique.

In {\bf State 1}, ads are visible on the page, either because the user has not installed an ad blocker or the ad blocker is ineffective on the page. The user could transition to {\bf State 2} by installing (or improving) her ad blocker, and in turn the publisher could return to {\bf State 1} by obfuscating their ads to bypass detection. This is one of three ``mini arms races'' that we will consider. A key question that we will address in Section \ref{sec:techniques} is whether these races could essentially go on forever or if there is a principled reason to expect that one side has a fundamental advantage.

From {\bf State 2}, the publisher could alternatively respond by escalating the arms race to the next level, namely by deploying ad blocking detection. Publishers have recently started turning to this technique. For surveys and measurement studies, see \cite{mughees2016first, nithyanand2016ad}. Once ad blocking is detected, publishers may nudge the user to disable it or block the user from accessing the site once it is disabled. We treat these two as identical; from a technical perspective, what matters is the successful detection of ad blocking (which defines {\bf State 3}).

Here the model already reveals its usefulness, in separating ad blocker circumvention into ad blocking detection and ad obfuscation. Current user-side tools generally do not consider this distinction and treat all ad blocker circumvention as a single problem to be solved with improved filter lists. However, this distinction is crucial because {\em robust} user-side countermeasures to the two techniques require very different approaches, as we will see in Section \ref{sec:techniques}.

The user could respond to ad blocking detection in two ways, the first of which is making ad blocking more stealthy (returning to {\bf State 2}). Stealth is different from obfuscation. Obfuscation seeks to make one type of content (or code) hard to distinguish from another, whereas stealth tries to hide the presence of some piece of code altogether. Thus, stealth is arguably a more powerful technique. In Section \ref{sec:techniques_stealth} we will see how root-kit style techniques can be applied to browser APIs to lie about the existence of an ad blocker. 

The second potential user response to ad blocking detection is what we call active ad blocking, which seeks to detect --- and then disable or modify --- any publisher code pertinent to ads. Whereas stealthy ad blocking seeks to avoid detection, active ad blocking instead seeks to interfere with the detector code, either by preventing proper detection from client-side code or preventing proper handling of detection from server-side code. Thus, in {\bf State 4}, both ads and ad blocking detection scripts are blocked. This may involve static techniques, such as searching for and modifying code ``signatures,'' or dynamic techniques, such as executing a shadow copy of the code to observe its behavior. In general, this is a hard problem. The distinction between stealth and active ad blocking is again one that has been missed by existing tools.

The publisher may obfuscate their ad blocking detection code to evade active ad blocking, returning to {\bf State 3}, and the user may in turn respond by improving the active ad blocking technique. But the model makes clear that no further escalation is possible. That is because active ad blocking runs at a higher privilege level than publisher code, so it cannot in turn be disabled --- at best it can be evaded. 

Thus, rather than the picture of eternal escalation evoked by the arms race metaphor, what we have instead is three ``mini arms races.'' In the rest of this paper, we will argue that rather than ad blocking being a game of whack-a-mole, there are principled ways for users and ad blockers to make progress over the state of the art in each of these three mini arms races.

%% file: newTechniques.tex
\section{New techniques and tools}\label{sec:techniques}
In this section we describe our new techniques for ad blocking and the proof-of-concept tools we have built. We start with a recap of traditional ad blocking before presenting our key concept, perceptual ad blocking. We defer the evaluation of these tools to Section \ref{sec:evaluation}. All three of the tools that we implemented (perceptual ad blocking, rootkit-style stealthy ad blocking, and signature-based active ad blocking) are available at: \anonymized{\url{https://github.com/futureAdblockingPaper/ad-blocking}}{\url{https://github.com/citp/ad-blocking}}.

\subsection{Traditional ad blocking}
\input{easylistTable}
Traditional ad blocking tools use {\em filter lists}, which are lists of known ad locations. The most prominent filter list, EasyList, is open source and maintained by a community of users. Enormous manual effort, however, goes into creating and maintaining the list; EasyList contains thousands of regular expressions.  Table \ref{tab:easylist_table} shows a snippet of rules from EasyList. Some of these rules refer to URLs and others refer to elements on web pages. 

When using ad blocking browser extensions such as Adblock Plus or uBlock Origin, the user subscribes to one or more filter lists. The extension periodically retrieves updated versions of the lists, and it applies the entire set of filters on every page visit. URL filters are applied to every outgoing request, and requests that match any filter are dropped. Element hiding filters are applied to the rendered web page, and elements that match any filter are hidden.

The set of features used in traditional adblocking is disjoint from the features that a human user might use to identify ads --- humans do not see URLs or HTML markup. One consequence is that traditional adblocking generalizes well to blocking of third-party trackers. Most trackers are not visible to the user, but filter lists can just as easily include regular expressions for tracker URLs as ad URLs. But a downside is that filter lists have to constantly play catch-up to stay in sync with the human categorization of ads.

\subsection{Perceptual ad blocking} \label{sec:techniques_perceptual}

Perceptual ad blocking seeks to improve resilience against ad obfuscation and minimize manual effort needed to create ad blockers. We rely on the key insight that ads are legally required to be clearly recognizable by humans. To make the method robust, we deliberately ignore all signals invisible to humans, including URLs and markup. Instead we consider visual and behavioral information. For example, an ad may include the text ``Sponsored'' or ``Close Ad'' within its boundaries, either directly or when hovered over.

We expect perceptual ad blocking to be less prone to an ``arms race.''
Ensuring that an advertisement serving system is compliant with legal and self-regulatory requirements takes time and expertise. Moreover, when multiple serving systems share  a self-regulatory compliance program, associated disclosures are usually implemented in a uniform and stable fashion. These properties--inertia and consolidation--are much more acture for perceptual ad blocking than traditional ad blocking. In order to defeat a filter list, all that is required is moving an advertisement to a different URL; in order to defeat a perceptual ad blocker, an entirely new ad disclosure standard must be approved. A filter list must account for diverse idiosyncracies in ad URLs; a perceptual ad blocker need only handle the small number of ad disclosure standards. There appears to be a small, stable group of such standards and implementations (Section \ref{sec:evaluation}). We discuss two such standards below.

We note, as an aside, that despite forceful FTC guidance about what constitutes clear disclosure, large minorities of consumers are unable to accurately distinguish online ads and most consumers are unfamiliar with AdChoices, by far the leading disclosure standard \cite{leon2012online, cranor-creepy}. These shortcomings for ordinary consumers do not pose a problem for perceptual ad blocking, because an expert system can be configured to detect cues regardless of general consumer familiarity. Our claim is that as long as the disclosure standards are unambiguous and adhered to, a perceptual ad blocker will have a 100\% recall at identifying ads governed by that standard. Indeed, our techniques could be seen as a way to highlight ads instead of block them, serving to enhance disclosure.\footnote{Taking this idea to this extreme, we may imagine a paradigm where websites merely mark up ads, and leave disclosure entirely to the user agent (browser). Different users may require different levels of visual cues to comfortably recognize ads.}

\begin{figure}
  \includegraphics[width=\linewidth]{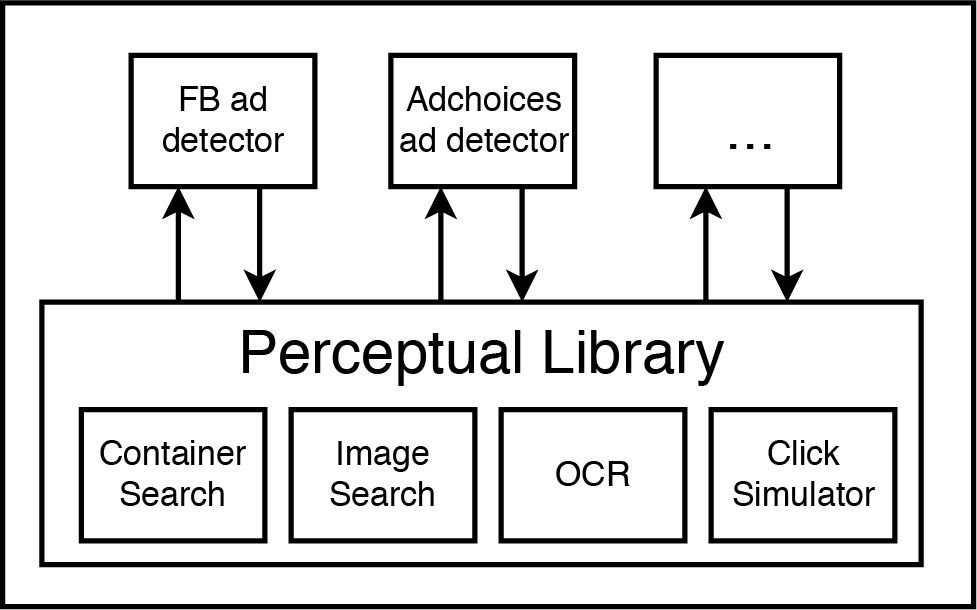}
  \caption{Architecture of perceptual ad blocking. Most of the ``heavy lifting'' happens in the perceptual library, allowing the ad blocker to quickly adapt to new ad disclosure standards.}
  \label{fig:perceptualDiagram}
\end{figure}

{\bf Design.}
Figure \ref{fig:perceptualDiagram} shows the high-level design of our perceptual ad blocker. It consists of a library of perceptual functionality which can be utilized by modules corresponding to various disclosure standards. We now describe the modules. The first is used to look for parts of the page that might potentially contain ads, and the latter three are used to actually detect the presence of ads.

\textbf{Container search API.} This allows querying for specific parts of the page based on perceptual features. It is similar to the standard DOM API functions getElementById and getElementsByClassName, except that instead of markup such as ID or class name, it accepts visual features. The list of visual features in our current prototype includes element type (e.g., link vs. button vs. text), size range, location on page, color; the client may also specify any other CSS properties. It also contains some heuristic intelligence and allows queries such as ``find sidebars''. 

\textbf{Fuzzy image matching and searching.} Disclosure standards may contain icons so as to be recognizable to the user, so perceptual ad blockers need the ability to search for such images in an ad. Exact comparison is not sufficient since such icons may have minor variations. So this API allows scanning for a particular image in any given container.

\textbf{Optical character recognition.} Some disclosures contain text rendered as images. Current instances of this appear to be due to poor programming practices, but it is also a technique that a publisher can use to attempt to obfuscate ads. Therefore optical character recognition (OCR) is an important component of a perceptual ad blocker. We use the Tesseract library\cite{ocrad}, which is a JavaScript OCR library. 

\textbf{Click simulator.} Finally, perceptual ad blocking also involves behavioral features, of which the main one is determining whether a click leads to a target link of interest (such as an ad disclosure statement page). Our click simulator API accepts either an element or (x, y) coordinate as input, simulates the click, opening any links in hidden tabs if applicable, follows redirects if necessary, and returns the destination URL. For an evaluation of these APIs, see Section \ref{sec:evaluation}. 

Together, these APIs allow specific perceptual ad blockers to be expressed succinctly. For example, our Facebook ad blocker below is expressed with logic similar to: {\em ``Retrieve all containers that are 450-550 pixels wide and have right and left borders (newsfeed ads) or are 225-325 pixels wide and inside a sidebar (sidebar ads). For each, look for the "sponsored" text and a link that leads to Facebook's ad disclosure page.''}

\textbf{Case study: AdChoices.} AdChoices is a self-regulatory program for behavioral advertising in North America and Europe. It is aimed at giving users more transparency and control over ads. It is recognizable by its icon (see Figure \ref{fig:adchoices}) which is shown on ads by participating companies, usually in the top right corner of the ad. The icon links to a page with more information about the ad or the website's collection practices. Despite the partial coverage of AdChoices in terms of geography, companies, and type of advertising, we find that over 60\% of ads in a sample of 183 ads from top news websites are covered by AdChoices (Section \ref{sec:evaluation}). 


\textbf{Case study: Facebook} A few large publishers have their own ad delivery systems, which may have their own disclosure standards, so perceptual ad blockers may need to treat them individually. Of these publishers, Facebook stands out for several reasons: it is the top website by traffic and time spent \cite{fb-traffic, fb-time-spent}, it has a complex ad targeting system, and it has recently made a concerted effort to obfuscate its ads.

On August 9 2016, Facebook announced a change to their site that ``renders all ad blockers useless'' \cite{nytimes-fb-adblocking}. The change in question was ad obfuscation: making the markup of ads sufficiently similar to that of regular Newsfeed posts that the two could not be distinguished by filter-list-based ad blockers. A few days later, Adblock Plus (EasyList) updated their filter rules, exploiting the fact that the obfuscation was not perfect, but Facebook in turn updated their markup. As of this writing, Adblock Plus remains unable to block Newsfeed ads on Facebook. This move to defeat ad blockers had a major financial impact for Facebook's bottom line, contributing to a 22\% growth (approximately USD 1 bn/year) in desktop advertisement revenue despite declining desktop usage \cite{facebook-ad-growth}.

EasyList targets elements to hide using CSS selectors. This allows extensions like Adblock Plus to block ads using only a CSS stylesheet, but severely limits the selection capability. A CSS selector allows selection of an element based on its own attributes and the attributes of all nodes containing it (e.g. ``div\#newsfeed div.ad'' would select divs with the class ``ad'' inside the div with the id ``newsfeed''), but not based on any of the elements that it contains. Facebook modified their page's DOM structure so that all Newsfeed item containers look identical whether they are ads or non-ads;\footnote{Any attributes that differ between an ad and non-ad, like ``id'', also differ between distinct non-ads.} any differences occur within the container and thus cannot be utilized by a CSS selector-based technique. 
\begin{figure}
  \includegraphics[width=\linewidth]{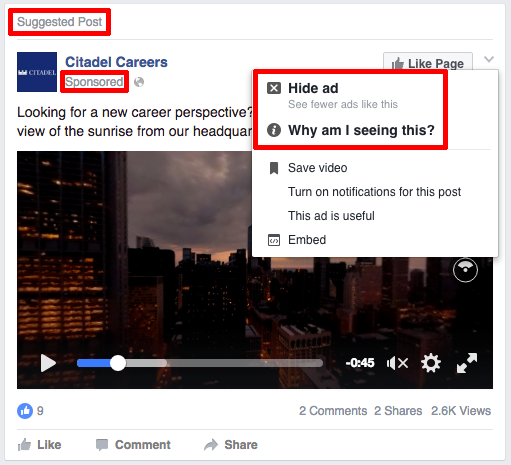}
  \caption{An example Facebook ad with disclosure highlighted}
  \label{fig:facebookAd}
\end{figure}

In contrast, our perceptual ad blocker does not rely on markup. Perceptual ad blocking is especially suited to Facebook, since the site has a strong disclosure standard (Figure \ref{fig:facebookAd}).
This might be due to the level of scrutiny of their advertising practices \cite{propublica}. 

{\em Technique.} First, we detect containers that might be ads, and for each such container, we run ad detection. Container detection works simply on the basis of the approximate physical dimension of the element. Of course, this could be circumvented by changing the dimensions, but that would impact the user experience.

To detect which containers correspond to ads, we detect the ``Sponsored'' text and the link. There are other possible markers such as dropdown menu which we don't currently utilize. The tool is configured to treat an item as an ad if any of the markers are found. Other configurations are possible, such as requiring all of the markers to be present.


\textbf{Comparison with filter list approach.} Perceptual ad blocking can potentially lead to a major reduction in development effort. Each perceptual module requires only a few tens of lines of code (Section \ref{sec:evaluation}). There are currently only a small number of disclosure standards on the web. However, if there is a proliferation of publisher-specific disclosure standards, this may erode the advantage of perceptual ad blocking.

Perceptual ad blocking is also more resilient to obfuscation compared to filter lists: with the latter, publishers need only change the URLs of ads and scripts to evade filters. 
Since it is a new technique, the resilience of perceptual ad blocking to adversarial evasion remains to be tested in practice (we discuss it briefly below). At the very least, it can complement the filter list approach since it works in a very different way. For example, perceptual ad detection could be used as part of web crawls to identify false positives and negatives of filter lists.

Finally, perceptual ad blocking can potentially reduce accidental breakage of sites caused by insufficiently specific patterns used in filter lists. Anecdotal evidence suggests that this is a widespread problem \cite{abp_forums}. However, the only rigorous evaluation of breakage caused by different approaches requires a large-scale crowdsourced study \cite{yu2016tracking}, so a rigorous comparison of breakage remains a topic for future work.

{\bf Our tool.} We released our perceptual ad blocker, with both the Adchoices and Facebook components, as a Chrome extension in mid-April 2017.\anonymized{[Link Redacted]}{\footnote{Our tool can be found at \url{https://chrome.google.com/webstore/detail/perceptual-ad-highlighter/mahgiflleahghaapkboihnbhdplhnchp}}}
Since it is a research prototype, we chose to sidestep potential ethical and legal concerns about ad blocking by merely highlighting ads rather than blocking them. 
As of this writing, our tool continues to be effective and has about 30,000 users on the Chrome web store. 

\textbf{Attacks on perceptual ad blocking and defenses.} Publishers may obfuscate ads in various ways to prevent perceptual ad blockers from detecting ads (false negatives), or cause them to detect non-ads as ads (false positives), or cause other kinds of collateral damage. We consider three main groups of attacks and defenses against them. We limit ourselves to ad obfuscation and assume that the publisher is unable to detect the existence of the ad blocker. Ad blocking detection will be considered in the next subsection.

The first attack is simply to change the visual display of ads. For example, our Facebook perceptual ad blocker looks for items of width between 450 and 550 pixels, so Facebook could pick a width outside this range. Developers of a filter list can push out updates just as quickly as a publisher can, so this could give rise to a ``whack-a-mole'' contest between a publisher and an ad blocker. This type of interaction, though, would require a publisher to deploy distracting changes to {\em all} users on a continual basis (recall that we are discussing ad obfuscation rather than ad blocking detection, so the publisher is unable to deploy this change to only users who employ ad blocking).

The second attack borrows from CAPTCHAs and seeks to obscure the visual presentation of the ad disclosure in a way that humans can understand but ad blockers cannot. Possible techniques include rendering ad disclosure text (such as ``Sponsored'') as a noisy image and obfuscating which elements are actually visible. As a simple example, both ad and non-ad elements could contain text such as ``Sponsored'', but it would be greyed out in non-ad elements. Our technique already incorporates defenses against this, such as fuzzy image matching and OCR. More importantly, these come at usability costs to users. For example, the use of images in place of text impedes the ability of visually impaired users to use screen readers.

The third attack is behavioral obfuscation. Perceptual ad blocking relies in part on determining the destination of links, which is not a visual feature but a behavioral one. There are various ways to obfuscate the existence of a link, the simplest of which are redirect chains and the use of JavaScript onClick events to handle clicks instead of the more standard href attribute. At worst, these and similar techniques will cause a performance degradation of the ad blocker due to the need to simulate clicking on an element rather than simply scan for the presence of links. A more devious variant of this attack is to obfuscate the difference between ad disclosure pages and ad landing pages. This attempts to force the perceptual ad blocker to click on ads, allowing the publisher to argue that its use constitutes click fraud. One defense against this is only simulate a link click after another ad disclosure marker (such as an AdChoices icon) has already been detected.

Finally, if perceptual ad blocking is successful, publishers and advertising intermediaries might attempt to evade legal requirements about advertising disclosure by relocating to more lenient jurisdictions. While we cannot rule out this possibility, we note that it is exceedingly risky for publishers and intermediaries. Regulatory requirements usually apply in the location where business is conducted, and not just the location where a firm is incorporated.  For example, the U.S. FTC (and state attorneys general) can enforce against foreign entities that do business in the United States. It can also hold domestic businesses, such as publishers, secondarily liable for violations by foreign partners, such as advertising networks. 

A fruitful area for future work would be to make perceptual ad blocking less dependent on ad disclosures. More powerful computer vision and AI techniques may enable detecting ads whether or not they are explicitly disclosed. We also present a policy recommendation:  improving the enforcement of misleading advertising laws is a light-touch way to enable the development of more robust ad blockers, thus improving web security and consumer protection.

 
\subsection{Ad blocking detection and stealth} \label{sec:techniques_stealth}
Ad blocking detection is evolving quickly. For surveys, see \cite{mughees2016first, nithyanand2016ad}. While the details are in flux, at a conceptual level we identify three main ways to detect ad blocking, broadly in line with previous work \cite{nithyanand2016ad}. (1) Detect the absence of known ads and scripts. This requires the publisher to hard-code the location of a ``well-known'' ad or (more typically) ad-related script such as DoubleClick. (2) Detect the absence of ``bait'' ads. The publisher inserts a ``fake'' ad and checks if it gets blocked. (3) Test side channels such as timing effects introduced by the ad blocking code. 

Ad blocking detection is arguably a more powerful technique than ad obfuscation, as it gives the publisher a variety of responses ranging from a mild appeal to the user to turn off ad blocking to outright disabling of content for ad block users. A more sophisticated response would be to allow only paying users to block ads.

In response to ad blocking detection, we propose and prototype the concept of stealthy ad blocking. As opposed to code obfuscation,\footnote{Our use of the term here and throughout the paper is very different from cryptographic obfuscation.} we seek principled and robust ways to hide ad blocking. As long as ad blocking is done locally (as opposed to blocking resources, which is the primary approach used by filter lists), we can hope that the publisher server observes no difference due to ad blocking, leaving us with only the task of fooling client-side JavaScript.

Defeating client-side checks is possible in principle because (in our model) the browser acts on behalf of the user and allows arbitrary customization. There are limits in practice, but extensions today do execute at a higher privilege level than publisher JavaScript code. Broadly, we identify two approaches to stealth. The first is to enumerate all the entry points in the browser API that the publisher code can use to interrogate the state of the page, and to fake the responses to each such query. This is exactly analogous to the technique used by rootkits at an OS level to hide their existence and activities. The second is to create two copies of the page, one which the user sees (and to which ad blocking will be applied) and one which the publisher code interacts with, and to ensure that information propagates between these copies in one direction but not the other.

\textbf{Rootkit-style stealthy ad blocking.} We present a general four-step technique for creating a browser ``rootkit'' and apply it to stealthy ad blocking. Our technique works by exploiting the separation between extension JavaScript context and page JavaScript context. This is a security-enhancing isolation feature intended to stop attacks such  as webpage scripts sniffing sensitive data from extensions. Our technique also exploits the fact that all API calls used by publisher JavaScript to examine the state of the page can be intercepted and modified by a browser extension by writing wrapper functions. 



Our approach to stealth works in conjunction with any existing technique for identifying web page elements that correspond to ads. Perceptual ad blocking falls into this category, as do  the ``element hiding" rules in existing filter lists, but not the resource blocking rules. Thus, our stealth approach is complementary to perceptual ad blocking (and is not dependent on it). Once ad containers are identified, the strategy is to visually block them using ``whitespace'' overlay containers, applying the rootkitting technique to prevent detection of these added containers. The details are as follows.

\begin{figure}
  \includegraphics[width=\linewidth]{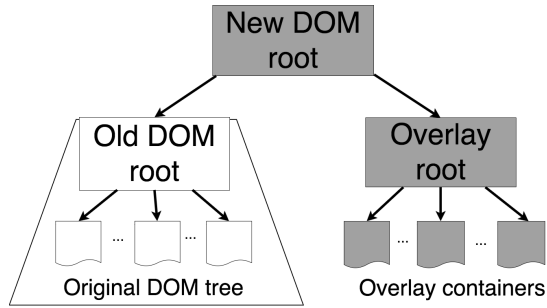}
  \caption{An illustration of DOM manipulations that enable stealthy ad blocking.}
  \label{fig:rootkit}
\end{figure}

\begin{enumerate}

\item {\em Add new DOM elements.} Figure \ref{fig:rootkit} shows how we add new DOM elements. Note that all overlay containers are in their own subtree rather than the more natural approach of making them sibling elements of the corresponding ads. This is necessary to hide their existence. Specifically, we replace all publisher JavaScript objects that could give away the existence of the new DOM root with versions that refer to the original DOM root. 

\item {\em Modify DOM traversal API.} We identify all API calls that can be used for traversing the DOM tree and replace them with modified versions so that publisher JavaScript cannot reach the actual DOM root or the subtree of overlay containers. This ensures that from the point of view of publisher JavaScript, the page appears unmodified. We identified and modified 40 such functions in the Mozilla JavaScript API (see Appendix \ref{sec:appendix_one} for details).

\item {\em Apply CSS rule addition and substitution.} We use a series of regular expressions to modify the original style sheets provided by the publisher so that the rules apply to the publisher subtree and not to the overlay subtree. We also add our own rules so that the overlay elements are placed in the right positions on the page. 

\item {\em Handle dynamic changes.} Using callbacks or timeouts, we monitor changes to the publisher subtree (e.g., a change in the size or position of an ad) that might necessitate changes to the overlay elements, and make those changes.

\end{enumerate}


\textbf{Attacks and defenses on rootkit-style stealthy ad blocking.} In general, rootkits are hard to get right in practice because of the need to anticipate every single API call that needs to be interfered with and because APIs change over time. 

Step 2 of our rootkitting technique (modifying the DOM traversal API) requires converting some JavaScript properties into functions, since {\tt element.parent} and other such accesses are properties and not functions. There are two ways to achieve this. The first, which we implemented, is overwriting these properties using JavaScript. The downside to this method is that publisher JavaScript can detect the interception of property access using the Object.getOwnPropertyDescriptor function on the object and property intercepted. Thus, the publisher will potentially discover that {\em some} modification to the code by {\em some} extension has taken place, but cannot discover what the changes are (see Appendix \ref{app:stealth} for details). Since there are a variety of reasons for extensions to intercept property access, the publisher cannot attribute the changes to ad blocking, and thus this method still foils ad blocking detection. The second method requires browser source-code modifications (Appendix \ref{sec:appendix_one}), but avoids this leakage. 



\textbf{Shadow execution style stealthy ad blocker.} We investigated an alternative approach to stealth that works as follows.

\begin{enumerate}
\item Create a shadow copy of the DOM tree.  Create a mapping from each element in the shadow copy to the corresponding element in the original.
\item Apply ad blocking to the original copy. 
\item Execute every API call on both copies --- in the original copy, map inputs to their corresponding elements in the unmodified shadow copy. Return responses only from the unmodified copy.
\end{enumerate}

This solution elegantly prevents publisher code from learning about the existence of ad blocking, since all responses returned to it come from the unmodified shadow copy of the DOM. It avoids the need to identify API functions that involve writes, because shadow execution is applied to {\em every} API call. Yet it ensures that all page functionality will be applied to the original copy which the user sees. 

Compared to the rootkit approach, the shadow approach has two advantages: it avoids the need to enumerate and individually modify all API calls that can be used for DOM traversal, and it also works with any ad blocker as a black box. The engineering effort required is massive, however, and requires significant modifications to browser source code. A further weakness is that the creation of two copies of the DOM may be detectable by the server if it did not properly avoid sending two sets of network requests. We regard this method to be primarily of conceptual interest, and did not pursue an implementation. For completeness, we analyze it as part of our systematization (Section \ref{sec:systematization}).

\subsection{Active ad blocking}
A few existing tools attempt to actively disable or combat ad blocking detection as opposed to the stealth approach of hiding the existence of ad blocking. For example, Adblock Plus has a feature to block pop-up messages that nudge the user to turn off their ad blocker. This does not affect ad blocking {\em detection}, however, just the action taken by the publisher. So in our model it does not correspond to a state transition --- in effect it is a ``no op.'' We regard this as a desirable feature of our model. Blocking popup nudges is only minimally effective, as a more adversarial publisher could simply disable the downloading and display of article content if ad blocking is detected.

The approach taken by the ``Anti-Adblock Killer'' tool \cite{antiadblockkiller} is somewhat more effective. To disable ad blocking detection, it blocks ad blocking detection scripts from being loaded and injects innocuous versions of the same code so that calls to the blocking detection script indicate there is no blocking. But this can be easily defeated. For example, on forbes.com the ad blocking detection script is wrapped in an anonymous function, and all JavaScript code is served from a single file, which together defeats the above techniques. 

Thus, more powerful methods are necessary. Effective active ad blocking requires the ability to modify JavaScript code. Some browsers expose the ability for extensions to modify HTTP response bodies (including scripts), but others don't, and a man-in-the-middle proxy is required.  Incidentally, this makes clear that active ad blocking is the final stage in the escalation of the ``arms race.'' Code that executes in a browser extension or in a proxy cannot in turn be overwritten. Thus, the publisher's response to active ad blocking can only be obfuscation, not further escalation. 

\junk{need to modify this section too. My current proposal:\\
We already have 1) current namespace overwriter\\ 
2) Describe the signature based, including the fact that we currently have the regex version working and can extend it to larger parsing [But grant still has to get this working] 3) Say that maintaining this list will also lead to analysis tools that could be used to run the differential ad blocker (again presented in the systematization because it is fundamentally different); but, also note the timing difficulties with such a tool and the fact it is not currently implementable.}

\textbf{Signature-based active ad blocking.} The simplest approach to active ad blocking is similar to the traditional filter list approach to ad blocking. The human analyst studies the publisher code on sites known to have ad blocking detectors, and constructs ``signatures'' or templates in a suitable abstraction. These signatures are shipped to the ad blocker which uses them to scan for known code in network traffic. The analyst also specifies how the code should be modified once found. Examples include changing the return value of a function, disabling a function altogether, or changing the value of a variable.

The simplest approach is string matching or regular expression matching, analogous to filter lists or byte signatures for malware. Like filter lists, it has obvious limitations, starting with randomization of function and variable names.  A more robust approach is structural JavaScript signatures, which is an ongoing area of research \cite{soni2015sicilian}. In this approach, the signature of a piece of code is invariant to a class of code transformations. In the malware domain, semantic signatures using call graphs have also been studied \cite{lee2010detecting}, but this technique has not been ported to JavaScript.

Our current implementation is based on regular expression matching, which is  sufficient for disabling all known ad blocking detectors so far (Section \ref{sec:evaluation}). There is a Firefox version that is a browser extension, as well as a Chrome version that requires mitmproxy, as it is not currently implementable as an extension in browsers other than Firefox.


\input{systematizationTable}

\textbf{Differential active ad blocking.} We outline a direction toward automating active ad blocking based on {\em differential execution} of the publisher code with and without ad blocking enabled. This is an open research problem. In fact, even in the domain of malware analysis which has received far more attention, comparable problems are not quite yet solvable using state of the art tools. For completeness, we analyze it as part of our systematization (Section \ref{sec:systematization}).


The technique works in four steps. (1) Execute the page with and without ad blocking enabled, carefully controlling all environment variables. (2) Use ``visual diffing'' to identify any impact of ad blocking detection on the ad blocked copy (e.g., a popup nudge to disable ad blocking). 
(3) Align execution traces and identify the code path that led to the visual difference potentially via techniques such as differential slicing \cite{johnson2011differential}.
(4) Identify API calls made by the publisher code along this set of code paths; replace their responses in the ad blocked version with the corresponding responses in the unmodified version.


%% file: easylistTable.tex
\begin{table*}[t]

\centering
\begin{tabular}{l l}
{\bf Rule} &
{\bf Rule Explanation}
    \\ \hline \\[-.7em]

\#\#\#Ad3Right
& Hides any container with the ID ``Ad3Right''
\\ \hline \\[-.7em]

liverpoolfc.com\#\#\#FooterLogos
& Hides any container with the ID ``FooterLogos'' on liverpoolfc.com
\\ \hline \\[-.7em]

$|$$|$atdhe.ws/pp.js
& Blocks any request to http(s)://atdhe.ws/pp.js
\\ \hline \\[-.7em]

.com/doubleclick/
& Blocks any request containing the substring ``.com/doubleclick/''

\\ \hline \\[-.3em]

\end{tabular}

\caption{\label{tab:easylist_table} Examples of EasyList filters --- the first two are ``element hiding'' and the latter two are ``resource blocking''. It is easy to see how ads can trivially be obfuscated to bypass such filters.}
\end{table*}

%% file: systematizationTable.tex
\begin{table*}[!ht]

\centering
\begin{tabular}{l C{1cm} C{1cm} C{1cm} C{1cm} C{1cm} C{1cm} C{1cm} C{1cm}}
Tool &
\rot{\footnotesize{\parbox{3cm}{\raggedright State transition\\
(Figure \ref{fig:statespace})}}} &
\rot{\footnotesize{\parbox{3cm}{\raggedright Blocks tracking}}} &

\rot{\footnotesize{\parbox{3.5cm}{\raggedright Blocks new ads/ad blocker detectors by default}}} &
\rot{\footnotesize{\parbox{3cm}{\raggedright Resilient to markup or code obfuscation}}} &
\rot{\footnotesize{\parbox{3.5cm}{\raggedright Cannot be detected by client-side scripts}}} &
\rot{\footnotesize{\parbox{3cm}{\raggedright Cannot be detected by server}}} &
\rot{\footnotesize{\parbox{3cm}{\raggedright Currently implementable as a browser extension}}} &
\rot{\footnotesize{\parbox{3cm}{\raggedright Order of growth of programming effort}}} 
    \\ \hline \\[-.7em]
Filter list 
&$1 \rightarrow 2$&$\CIRCLE$&X&X&X&X&$\CIRCLE$ &F+T    
\\ \hline \\[-.7em]

{\bf Perceptual}
&$1 \rightarrow 2$&X&$\CIRCLE$&$\CIRCLE$&X&$\LEFTcircle$&$\CIRCLE$&D    
\\ \hline  \\[-.7em]

{\bf Rootkit-style}
&$3 \rightarrow 2$&X&N/A&N/A&$\LEFTcircle$&$\CIRCLE$&$\CIRCLE$&B   
\\ \hline \\[-.7em]

{\em Shadow copy-based}
&$3 \rightarrow 2$&X&N/A&N/A&$\CIRCLE$&$\LEFTcircle$&X&1    
\\ \hline  \\[-.7em]

Namespace overwriter
&$3 \rightarrow 4$&X&X&X&X&X&$\CIRCLE$&S    
\\ \hline \\[-.7em]

{\bf Signature-based}
&$3 \rightarrow 4$&Maybe&X&X&X&$\CIRCLE$&\LEFTcircle&S    
\\ \hline \\[-.7em]

{\em Differential execution-based}
&$3 \rightarrow 4$&X&$\CIRCLE$&$\CIRCLE$&\CIRCLE&$\LEFTcircle$&X&1    
\\ \hline \\[-.3em]

\end{tabular}

\caption{\label{tab:tools_systematization_table} Systematization of tools. A half filled-circle means partial achievement of goal. For the order of growth, F = \# of 1st party site; T = \# 3rd party advertising scripts, D = \# of disclosure standards, S = \# of Ad blocker detector scripts, B = size of browser API. Tools in regular font are existing ones; those in {\bf bold} are proposed,  implemented, and evaluated by us, and the ones in {\em italics} are possible designs that we include here for completeness of analyzing the design space.} 
\end{table*}

%% file: systematization.tex
\section{Systematization and evaluation}
\label{sec:systematization}

\subsection{Systematization}

Table \ref{tab:tools_systematization_table} systematizes all the techniques described in Section \ref{sec:techniques}. We define and discuss each of the properties (columns), highlighting noteworthy differences between the techniques. 

\textbf{Blocks tracking}: evaluates whether the technique prevents tracking of users. Tracker blocking is inherently at odds with the idea of stealth, since it requires blocking network requests, which is detectable by the server. Active ad blocking techniques are also not designed to prevent tracking, though the signature-based active ad blocker can potentially be configured to prevent tracking and tracker blocking detection.

\textbf{Blocks new ads/ad block detectors by default}: describes whether new ads (in the case of ad blockers) or ad blocking detectors (in the case of active ad blockers) are blocked by default, rather than requiring manual updating. For example, the differential execution-based active ad blocker would automatically disable a novel ad block detection technique on a site that had not employed one previously, but our signature-based active ad blocker would require adding a new site/signature pair in order to disable the detector. For the two stealth techniques (rootkit-style and shadow copy-based), the behavior is determined by the underlying ad detection method on which they are built.

\textbf{Resilient to markup or code obfuscation}: describes whether the technique is resistant to obfuscation of markup in existing ads, or to obfuscation of code in existing ad blocking detectors. For example, changing the ID of an ad container would defeat a filter in a filter list-based ad blocker, but it would have no effect on the perceptual ad blocker as long as the ad consisted of the same visual features. Again, for the stealthy ad blockers, this behavior depends on the underlying ad detection method. 

\textbf{Cannot be detected by client-side script}: this is the definition of stealth. When our rootkit-style stealthy ad blocker is in use, the adversary can detect that there have been some changes, but cannot determine whether ads are being blocked or not (see Appendix \ref{app:stealth} for details). As for the active ad blockers, namespace overwriting and signature-based techniques work by manipulating either the DOM or script bodies, and are hence detectable. In contrast, the differential ad blocking technique modifies {\em all} code that leads to a different output when ads are blocked, so any client-side scripts that attempt to detect it will also be interfered with.

\textbf{Cannot be detected by the server}: describes whether the use of the technique affects the network traffic to the server, allowing the server to detect its use. This is another key property of stealth and a desirable requirement for active ad blockers. Both filter list-based ad blockers and the namespace overwriter active ad blocker block requests to the server, which is necessarily detectable by the server. The optional link-click simulator module of the perceptual ad blocker is detectable by the server, but the other modules are not. If the shadow copy-based stealthy ad blocker and the differential execution-based active ad blocker do not carefully handle requests to prevent duplicate requests from reaching the server, the server may be able to detect ad blocking because of the duplicate requests. 

\textbf{Currently implementable as a browser extension}: describes whether this can be currently implemented as a browser extension. We have implemented rootkit-style stealthy ad blocker as an extension but it would require browser modification to prevent the adversary from detecting any modification to the page. The signature-based active ad blocker is currently implementable as a browser extension in Firefox, but requires a proxy in other browsers.

\textbf{Order of growth of programming effort}: evaluates the programming effort associated with these tools. For the filter-list based ad blocker, resource blocking rules operate on a per-third-party basis and element hiding rules operate on a per-first-party basis. The programming effort for the perceptual adblocker is independent of the number of first parties and third parties, and only depends on the number of disclosure standards, although the ``hidden constant'' --- that is, the amount of work per standard --- is higher than for the filter list. For stealth ad blockers, the programming effort to achieve stealth for the rootkit-style is based on the size of the browser API and the effort for the shadow copy-based stealth ad blocker, while considerable, is not dependent on any of the factors studied in this column. Both the namespace overwriter and signature-based active adblockers have work proportional to the number of ad blocker detector scripts, with an especially large ``hidden constant'' (Section 5.2). Like the shadow copy-based stealth ad blocker, the programming effort for the differential execution-based active ad blocker is a large constant but does not depend on the number of detector scripts in existence.

\subsection{Evaluation}\label{sec:evaluation}

Here we evaluate the effectiveness of each of the three techniques that we have prototyped, along with the filter list technique. We defer performance measurements to Appendix \ref{app:performance}, and emphasize that our focus in this paper is explicitly on the long-term evolution of this space. We do {\em not} claim that all our prototypes have acceptable performance on commodity machines today (although most do). However, we expect this to change over time both due to Moore's law and due to the gradual performance improvement of JavaScript relative to native code.

{\bf Filter list ad blocker.} We argued earlier that creating and maintaining filter lists is a laborious process. To verify this, we downloaded the version history of EasyList, the primary general-purpose ad blocking filter list in use. It has had $\sim$8,000 git commits in 2015, an average of $\sim$660 per month. These commits added $\sim$21,600 filters and removed $\sim$10,000. Over the history of the project, 12 authors have each made over 1,000 commits. 

\begin{table}[tbp]
\centering
\begin{tabular}{c|cc}

                    & \shortstack[c]{Reported as \\ Adchoices ad} & \shortstack[c]{Not reported as \\ Adchoices ad} \\ \hline \\[-5pt]
\shortstack[c]{Contains \\ Adchoices ad}         & 208                      & 4\\[5pt]                            
\shortstack[c]{Does not contain \\ Adchoices ad} & 3                        & 238                          \\ 
\end{tabular}
\caption{Evaluation of the Perceptual ad blocker in identifying Adchoices ads.}
\label{tab:adchoices_eval}
\end{table}

{\bf Perceptual ad blocker.} Recall that our perceptual ad blocker consists of a generic library of perceptual functionality, using which we implemented two specific perceptual ad blockers. We now evaluate the effectiveness of these two.

{\em Facebook.} We evaluated this extension on a sample of 50 Facebook ads, including 35 from the main newsfeed, 10 from the main page sidebar, and 5 from the sidebar of a ``trending topics'' page. All 50 ads were correctly identified. Further, the authors have observed no false positives or negatives in over six months of regular use.

{\em AdChoices.} We evaluated this extension a random sample of 100 sites from the Alexa top 500 news sites. To establish ground truth, we manually identified 453 iFrames, of which 212 contained AdChoices ads. In the vast majority of cases the extension correctly reported the presence or absence of an AdChoices ad, with only 3 false positives and 4 false negatives (Table \ref{tab:adchoices_eval}).

{\em AdChoices coverage on the web.} We examined on a random sample of 20 of the Alexa top 500 news websites, visited both the home page and one article page on each of these sites, and collected a total of 173 ads. We manually classified these to identify which disclosure standard (if any) was used. The results are summarized in Table~\ref{tab:adchoices}.

\input{adchoicesTable}

First, about a fifth of ads (37/173) contained no disclosure. These ads are potentially non-compliant with the law, and we expect that this fraction will go down over time as the industry matures. Of the remaining, the vast majority followed AdChoices, with a few different implementations as shown in the table. A small number (4) were Outbrain, a third-party content recommendation site with its own disclosure standard. The remainder contained no disclosure in the ad iFrame itself and instead contained a disclosure by the publisher surrounding the ad. The adoption of AdChoices is a tremendous increase from measurements conducted a few years ago \cite{komanduri2011adchoices}, and supports our view that ad disclosures are becoming more ubiquitous and standardized over time. 

{\em Succinctness.} Most of the complexity of the implementation is in the perceptual library (578 lines of JavaScript code). There is also a small library of utility functions (76 lines of code). The AdChoices ad blocker is only 28 lines of code, and the Facebook ad blocker is only 63 lines of code (it is longer since there are two types of Facebook ads, newsfeed ads and sidebar ads, with different perceptual properties). All the above counts ignore empty lines and comments.

{\bf Stealthy ad blocker.} 
We created a stealthy ad blocker based on EasyList. We chose EasyList to make our stealthy ad blocker independent of the perceptual ad blocker, given that the latter is not yet field tested (except on Facebook). We emphasize that stealth is orthogonal to the underlying ad blocking technique, as long as the technique is capable of returning a list of containers corresponding to ads on any given web page. 

We identified a list of 50 sites from the filter list of the ``Anti-Adblock Killer'' tool that were indicative of ad blocking detection. We manually verified the presence of ad blocking detection on these sites. We remove resource blocking rules from EasyList and add a few element hiding rules added to block all ads on these 50 sites. Next, we visit each of these sites with our stealth ad blocker and confirm that it is undetected while blocking ads on all 50 of those sites.

{\bf Signature-based active ad blocker.} We evaluated our signature-based active ad blocker on the same list of 50 sites in February 2017. 

A 2016 measurement study found a small number of implementations of ad blocking detection being shared and reused \cite{nithyanand2016ad}. In contrast, in early 2017 we found a radically different picture. There is a proliferation of implementations of ad blocking detection, and there were 40 different implementations among these 50 sites! We consider implementations to be different if they use different techniques for detecting ad blockers.\footnote{We have compiled a list of these techniques in an online supplement at \anonymized{\url{https://github.com/futureAdblockingPaper/ad-blocking/blob/master/signature-active/detectorImplementationsTable.pdf}}{\url{https://citp.github.io/ad-blocking/signature-active/detectorImplementationsTable.html}}} In fact, a few sites use a multitude of different techniques, the maximum being six.

For each of these 50 sites, we are able to create regular expression-based signatures of ad blocking detection. Thus, while further evolution of this arms race may necessitate more powerful approaches such as structural or semantic signatures, vanilla signatures are currently sufficient. Further, in all cases, simple code snippets suffice to disable ad blocking detection --- either forcing the return value of a function or removal of a ``bait'' element inserted into the DOM by the ad blocking detection script. 

Thus, signature-based active ad blocking is currently a viable approach, but it scales poorly due to the proliferation of implementations. We conjecture that the reason that most publishers have developed their own implementations is precisely to avoid getting blocked by existing active ad blockers. Differential active ad blocking, on the other hand, will require a long-term research effort. We conclude that stealthy ad blocking, rather than active ad blocking, offers users the best hope of gaining the upper hand in the arms race.

%% file: adchoicesTable.tex
\begin{table}[t]

\centering

\begin{tabular}{l l r}
\hline \\[-.7em]
\multicolumn{2}{l}{Disclosure} & \# of occurrences   \\ \hline \\[-.7em]
\multicolumn{2}{l}{Adchoices } &  \\
\multirow{ 5}{*}{ \hspace{5mm} }
& Google & 34     \\
& TRUSTe & 26     \\
& Ghostery & 21     \\
& Taboola & 8     \\
& Other & 18 \\
& Adchoices total & 107 \\
\multicolumn{2}{l}{Outbrain} & 4 \\
\multicolumn{2}{l}{Publisher Disclosure Only} & 25 \\
\multicolumn{2}{l}{No Disclosure} & 37 \\
\multicolumn{2}{l}{Total} & 173
\\ \hline \\[-.7em]
\end{tabular}

\caption{\label{tab:adchoices} Breakdown of a sample of analyzed ads by type of disclosure. }
\end{table}

%% file: legal.tex
\section{Legal analysis}
The prior sections have characterized the technical dimensions of the ``arms race'' between users and publishers. We considered legal constraints on publishers and advertisers as part of perceptual ad blocking, but there are also legal issues associated with ad blocking technology, which we analyze in this section. We conclude that the law does not alter our analysis of users' ability to deploy sophisticated ad blocking techniques. 

We address American law because the largest advertising and advertising-supported businesses are based in the United States, and because policy disputes about advertisement blocking have tended to arise in the United States \cite{randall-wsj}. Furthermore, if American law permits ad blocking (which we conclude that it generally does), there is no realistic mechanism for other jurisdictions to inhibit consumer adoption of ad blocking. We note that Adblock Plus has been a target of strategic litigation in Germany, and has so far prevailed under German law. 

We focus on legal liability for providers of advertisement blocking tools, rather than liability for users who block advertisements, because litigation against end users has not proven a successful strategy for addressing widespread online conduct. For example, the music industry abandoned its strategy of litigation against end users, owing to the high cost, limited benefits, and public backlash \cite{music-suits-wsj}.
 
{\bf Tortious Interference.} Users routinely form contracts with publishers, either through the act of visiting a website or by creating an account  \cite{lemley2006terms}. Even absent contract formation, publishers might reasonably expect economic advantage (i.e. advertising revenue) in conjunction with a user's access to content. State law claims sounding in tortious interference could enable publishers to vindicate these contractual or economic interests, and provide injunctive and monetary remedies against developers of ad blockers. Tortious interference has previously proven a viable cause of action in related contexts involving user-directed modification to software, such as litigation against developers of video game cheating software and businesses that offer support for enterprise software.\footnote{See MDY Industries, LLC v. Blizzard Entertainment, Inc., 629 F.3d 928 (9th Cir. 2010); Oracle America, Inc. v. Hewlett Packard Enterprise Co., No. 16-cv-01393-JST, 2016 WL 3951653 (N.D. Cal. July 22, 2016).}

A tortious interference claim would be unlikely to succeed against an ad blocker developer, however, for three reasons. First, a publisher would have to prove that the developer had knowledge of the contractual relationship or economic advantage. When a developer builds a general-purpose ad blocker, it lacks the requisite knowledge.\footnote{This defense is especially applicable to perceptual ad blocking (except when specific to a first-party website) and the other generic techniques that we develop in this paper. Custom filter rules, by contrast, could satisfy the knowledge requirement because they relate to blocking ads associated with specific publishers.} Second, a publisher would usually have to demonstrate that the developer intentionally interfered with its contracts or economic advantage. While it is true that ad blocking software has the effect of reducing ad revenue, its purpose is to benefit consumers with fewer disruptions. Third, tortious interference claims generally require proof of wrongdoing, such as misrepresentation or fraud. There is no plausible wrongdoing between the developer of an advertisement blocker and the users who install the blocker; on the contrary, the users have elected to install the blocker precisely because of how it operates.
 
{\bf Computer Abuse Law.}  Federal and state computer abuse statutes, most notably the Computer Fraud and Abuse Act of 1986, have at times been interpreted to encompass breaches of online contracts \cite{mayerNosal}. Even if that interpretation of computer abuse law were accurate --- most courts now reject it --- publishers would still not have a cudgel against developers of advertisement blockers. Computer abuse law does not generally provide a civil cause of action for secondary liability (i.e. causing someone else to violate the law).\footnote{See Netapp, Inc. v. Nimble Storage, Inc., 41 F. Supp. 3d 816 (N.D. Cal. 2014).} And while there is secondary criminal liability, a prosecutor would have to prove intent --- a hurdle which is, as discussed above, likely insurmountable. We are also exceedingly skeptical that a federal or state prosecutor would elect to bring criminal charges against the developer of an ad blocking tool.
 
{\bf Copyright Law.} Federal copyright law, unlike computer crime law, does provide for secondary civil liability.\footnote{See MGM Studios, Inc. v. Grokster, Ltd., 545 U.S. 913 (2005); A\&M Records, Inc. v. Napster, Inc., 239 F.3d 1004 (2001).} And, if a publisher were to prevail, it could recover up to \$150,000 per instance of infringement.\footnote{See 17 U.S.C. \S 504.} A copyright claim against an ad blocker developer would be unlikely to succeed, though, for the simple reason that ad blocking does not appear to involve duplicating a copyrighted work or creating a derivative work. And even if ad blocking were a copyright infringement, developers would have a strong defense of fair use.\footnote{See Sony Corp. of America v. Universal City Studios, Inc., 464 U.S. 417 (1984), the ``Betamax case.''}
 
{\bf Digital Millennium Copyright Act.} The anti-circumvention provisions of the Digital Millennium Copyright Act (DMCA) prohibit circumventing a technical protection that limits access to a copyrighted work.\footnote{See 17 U.S.C. \S 1201.} Unlike computer abuse law, the DMCA does include secondary liability that could reach software developers. But, critically, DMCA protections only apply if there is an effective technical protection in place. A trivial protection, such as a no-copy bit or obfuscation, is not sufficient to sustain a DMCA claim.\footnote{See MDY Industries, LLC v. Blizzard Entertainment, Inc., 629 F.3d 928 (9th Cir. 2010); Agfa Monotype Corp. v. Adobe Systems, Inc., 404 F. Supp. 2d 1030 (N.D. Ill. 2005).} So long as a publisher does not impose digital rights management (DRM) protection for its content ---- a practice that is rare today outside of audio and video content --- the DMCA does not provide a legal tool against ad blocker developers. In particular, current publisher protection mechanisms involve merely hiding article content with an overlay element if an ad blocker is detected, which would be considered a trivial protection.

%% file: appendices.tex
\clearpage
\appendix
\section{Appendices}
\subsection{List of Javascript functions and properties modified by the rootkit-style stealthy ad blocker}\label{sec:appendix_one}

\begin{itemize}

\item Properties on the document: 9
\begin{itemize}
\item  childNodes
\item  children
\item  documentElement
\item  firstElementChild
\item  lastChild
\item  lastElementChild
\item  body
\item  scrollingElement
\item  all
\end{itemize}

\item Functions on the document: 7
\begin{itemize}
\item  getElementById
\item  getElementsByClassName
\item  getElementsByTagName
\item  getElementsByTagNameNS
\item  querySelector
\item  querySelectorAll
\item  \textit{elementFromPoint} (requires the experimental elementsFromPoint function)
\end{itemize}

\item Properties on the fake HTML: 11
\begin{itemize}
\item  tagName
\item  nodeName
\item  localName
\item  parentNode
\item  parentElement
\item  firstChild
\item  firstElementChild
\item  childElementCount
\item  id
\item  outerHTML
\item  innerHTML
\end{itemize}

\item Functions on the fake HTML: 1
\begin{itemize}
\item  insertBefore
\end{itemize}

\item Properties on head: 4
\begin{itemize}
\item  nextElementSibling
\item  nextSibling
\item  parentElement
\item  parentNode
\end{itemize}

\item Properties on the fake body: 8
\begin{itemize}
\item  tagName
\item  nodeName
\item  localName
\item  previousElementSibling
\item  previousSibling
\item  id
\item  className
\item  outerHTML
\end{itemize}

\end{itemize}

\subsection{Performance}
\label{app:performance}
We study the long-term evolution of ad blocking. The performance numbers reported here may change dramatically over time, and should be taken in context.

We test the perceptual ad blocker and the rootkit-style stealthy ad blocker on a random sample of 10 sites from the Alexa top 500. We visited each site 5 times with a clean browser profile (i.e., having cleared cache and other local state). These tests were done on a Macbook Pro running OSX 10.12.2 with a 2.5ghz Intel Core i7 processor, 16 GB of memory, and a 85 Megabits/s network.

{\bf Perceptual ad blocker.} We found that the ad blocker adds $0.53 \pm 0.15$ seconds of latency to page load times. We disabled the OCR module for these measurements, as it is slow (about 1 second per image), as mentioned earlier. The perceptual ad blocker has nearly identical effectiveness on the sample of websites we tested even with this module disabled. We also note that OCR implemented using native code is likely to be much faster: we tested the Tesseract C++ implementation and found it to be about an order of magnitude faster than the JavaScript implementation.

{\bf Rootkit-style stealthy ad blocker.}  We used the firing of the {\tt DOMContentLoaded} event as a proxy for loading of the page. With no ad blocker, page loads take $1.4 \pm 0.27$ seconds. With an ad blocker, page loads take $1.47 \pm 0.1$ seconds, an overhead of 70 milliseconds per page load.

\subsection{Undetectability of stealthy ad blocking}
\label{app:stealth}
As discussed in Section \ref{sec:techniques_stealth}, the key step in creating a stealthy adblocker using a rootkit technique is to overwrite JavaScript functions that can be used for DOM traversal, like this:\begin{verbatim} 
   window.targetFunc = function() {
      return <modified_code>;}
\end{verbatim}

However, this modification can be both detected and inspected by an adversary. An adversary can call one of three source inspection methods: \texttt{toString()}, \texttt{toLocaleString()} (which just calls \texttt{toString()} by default) and \texttt{toSource()}. Using \texttt{toString} as an example, the adversary would call \texttt{window.targetFunc.toString()}, which provides a string copy of the function’s code; in this case,\begin{verbatim} 
   function() { return <modified_code>;}
\end{verbatim} 

Every function inherits this default \texttt{toString} method from the \texttt{function} prototype. To prevent inspection, the stealthy adblocker can overwrite the function's default \texttt{toString} with a modified version. For example: \begin{verbatim} 
   window.targetFunc.toString = function() {
      return <original_code>;}
\end{verbatim}

However, the adversary can circumvent this as well, by finding another function, say \texttt{window.innocentFunc}, and copying the \texttt{toString} function from \texttt{innocentFunc} to \texttt{targetFunc}. This ensures that \texttt{targetFunc.toString()} will return \begin{verbatim} 
   function() { return <modified_code>; }
\end{verbatim}

To prevent this, instead of modifying \texttt{toString} on a case-by-case basis, we modify the universal \texttt{toString} function on the prototype to hide all of our modifications.

Finally, the adversary may turn to {\em protected objects}. These are objects or properties that scripts, including those in extensions, cannot overwrite. For example, some security properties rely on the property \texttt{window.location} to be accurate, so JavaScript code cannot overwrite this property \cite{bugzillaLocationSpoofing}. Because of the adversary's ability to transfer any correct \texttt{toString} function to one of the modified functions, if any function's \texttt{toString} method is protected, it can be used to reveal the details of our modification. We found no protected objects that allowed this attack at the time of this writing, but if such a protected object was introduced in the future it will cause difficulties for our method, and might require a browser modification to fully protect against.